# Magnetic shape-memory effects in La$_{2-x}$Sr$_x$CuO$_4$ crystals

**A. N. Lavrov, Seiki Komiya & Yoichi Ando**

Central Research Institute of Electric Power Industry, Komae, Tokyo 201-8511, Japan

---------------------------------------------------------------

The magnetic field affects the motion of electrons and the orientation of spins in solids, but it is believed to have little impact on the crystal structure. This common perception has been challenged recently by ferromagnetic shape-memory alloys[1,2], where the spin-lattice coupling is so strong that crystallographic axes even in a *fixed* sample are forced to rotate, following the direction of moments. One would, however, least expect any structural change to be induced in antiferromagnets where spins are antiparallel and give no net moment. Here we report on such unexpected magnetic shape-memory effects that take place ironically in one of the best-studied 2D antiferromagnets[3], La$_{2-x}$Sr$_x$CuO$_4$ (LSCO). We find that lightly-doped LSCO crystals tend to align their *b* axis along the magnetic field, and if the crystal orientation is fixed, this alignment occurs through the generation and motion of crystallographic twin boundaries. Both resistivity and magnetic susceptibility exhibit curious switching and memory effects induced by the crystal-axes rotation; moreover, clear kinks moving over the crystal surfaces allow one to watch the crystal rearrangement directly with a microscope or even bare eyes.

However strong the presently available magnetic fields are, their influence on a single electron in a solid is still too weak to compete with thermal fluctuations at room temperature ($\mu_B H \ll k_B T$). Unless this weak magnetic influence is integrated over macroscopic number of electrons, as happens, for example, in ferromagnets, the magnetic field is incapable of inducing any sharp, qualitative effect at high temperatures. To our surprise, lightly doped La$_{2-x}$Sr$_x$CuO$_4$ – the prototype layered copper oxide – seems to flout this common-sense rule. At the doping level $x = 0.01$, for instance, LSCO has no static magnetic order at room temperature[3-5] (though Cu$^{2+}$ spins order antiferromagnetically below the Néel temperature $T_N \sim 200$ K; Refs. 4–6), nor has a static charge order[7,8]. Nevertheless, the magnetic field applied at around room temperature to such $x = 0.01$ samples (high-quality LSCO crystals were grown by the traveling-solvent floating-zone technique[4,6]) is able to generate intriguing memory effects in the in-plane resistivity $\rho_{ab}$. For example, in an experiment shown in Fig. 1a, upon first application of a strong in-plane magnetic field perpendicular to the current, $\rho_{ab}$ increases and never comes back to the initial value after switching off the field (Fig. 1b). Moreover, the effect is far from being isotropic: while the resistivity transverse to the magnetic field increases, it decreases in the direction along the field (Fig. 1c). By applying the magnetic field across or along the sample's axis, we can reversibly drive the resistivity up and down ($\Delta\rho_{ab} = 1-4\%$, depending on the sample and temperature), each of the states being stable enough to survive subsequent room-temperature storing for several days and temperature sweeping (Fig. 1d). Interestingly, only in samples cut along the orthorhombic *a* or *b* axes the resistivity exhibits memory effects, while in samples cut along the diagonal direction, or along the *c* axis, it does not – apparently, the magnetic field just switches the easier-conduction route between two orthorhombic domains.

The magnetic switching of the resistivity observed here is unprecedented for non-magnetically-ordered materials, and challenges us with questions not only how the magnetic field overcomes the effect of thermal fluctuations, but also what is actually changing in LSCO samples to affect their resistivity. The memory effect in LSCO certainly differs from that in ferromagnets, where magnetic domains rearrange with field. It does also differ from the memory effect[9] in another copper oxide YBa$_2$Cu$_3$O$_{6+x}$, which occurs at liquid-helium temperatures and is related to the freezing of antiferromagnetic domains and their charged boundaries[9]; in the present case the temperature is 100 times higher, and, moreover, the LSCO ($x = 0.01$) crystals studied here have $T_N$ of only 160–200 K and there is just no static magnetic order at room temperature to be altered. In the absence of any static magnetic or charge order, we inevitably arrive at an intriguing conclusion that it must be the crystal structure that is responsible for the memory effects.

The symmetry of the otherwise perfect square lattice of LSCO is broken by a weak (~1%) orthorhombic distortion, which emerges below the tetragonal-to-orthorhombic transition[3] ($T_{TO} = 400-500$ K). This distortion actually makes the resistivity slightly anisotropic, where $\rho_a$ exceeds $\rho_b$ by several percent at room temperature[8]. Given this resistivity anisotropy, the resistivity switching would be easily understood, if the magnetic field were capable of swapping the orientation of crystallographic *a* and *b* axes, so that the sample shrinks by 1% along one direction and expands along another. One might be inclined to abandon such a seemingly unrealistic hypothesis – LSCO has a magnetic susceptibility as small as that of plastic or paper[4], and even in such ferromagnetic metals as Fe, Ni, or Co the magnetic field can induce only a subtle (less than 0.01%) contraction or expansion of the lattice. Owing to its counter-intuitiveness, a possibility of the magneto-structural rearrangement was not even considered in this otherwise exceptionally well-studied[3] material. However, to our surprise, the x-ray measurements indeed confirm that the crystal axes in LSCO rotate with the magnetic field; moreover, we have accidentally discovered that the crystal shape so vividly varies with the crystallographic rearrangement that one can watch it directly with an optical microscope.

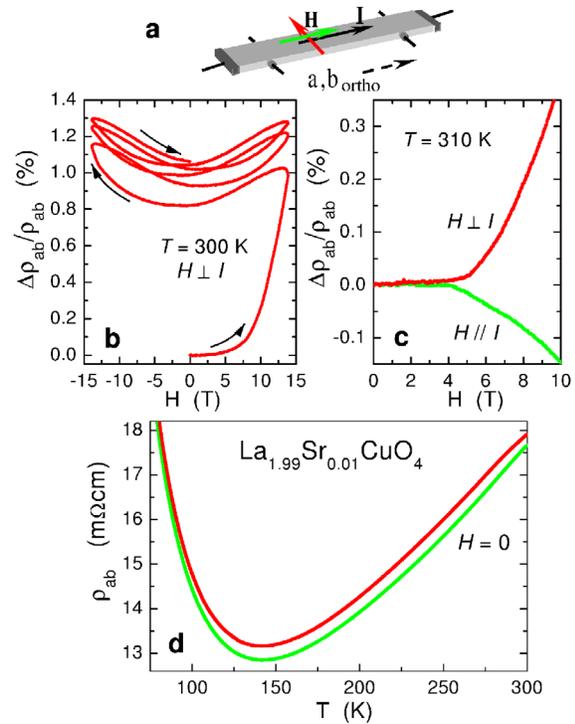

**Figure 1.** Temperature and magnetic-field dependence of the in-plane resistivity $\rho_{ab}$ demonstrating the memory effect in LSCO. **a**, Schematic of measurements: the sample is cut along the orthorhombic *a*(*b*) axis and the in-plane magnetic field is applied along or perpendicular to the current. **b**, Variation of $\rho_{ab}$ upon sweeping the magnetic field $H \perp I$. The first application of a strong magnetic field at room temperature induces an essentially irreversible change of $\rho_{ab}$ – neither switching off the field nor altering its polarity recovers the initial state. **c**, The resistivity increases with $H$ applied transverse to the current (red curve), while it decreases for $H \parallel I$ (green curve); the magnetic field acts like switching a traffic light for the current flow. The field-induced states are stable at least for several days and in **d** we show zero-field $\rho_{ab}(T)$ data measured after applying a 14-T field $H \perp I$ (red) and $H \parallel I$ (green) at room temperature.

When LSCO crystals are cooled through the tetragonal-orthorhombic (TO) transition, they usually develop a domain structure (twins), where the orthorhombic distortion alters its sign upon crossing the domain boundaries (Fig. 2a). Clear kinks generated by the domain boundaries appear on initially flat $ac/bc$ faces[8] and can be easily observed with an optical microscope, or even bare eyes. Recently we succeeded in detwinning LSCO crystals by cooling them through the TO transition under a uniaxial stress of 30–100 MPa, where the removal of domain boundaries was controlled by the optical microscopy[4,8]. In Figures 2b–2f, the same technique is employed to demonstrate the remarkable effect of the magnetic field – depending on the field direction, the crystal vividly changes, new waves emerge on the surface, move, and then disappear completely. Clearly, LSCO crystals try to align their $b$-axis along the magnetic field by generating and/or moving the domain boundaries. The particular specimen pictured in Fig. 2 actually becomes completely detwinned (Figs. 2d, 2f), as it does after application of a uniaxial stress, for one of the field directions.

For the other direction of the field, though, a perfect alignment is not achieved (Figs. 2c, 2e), presumably because of some internal crystal strain. While the crystal structure exhibits such a striking agility under the impact of the magnetic field, we have confirmed the detwinned state in Figs. 2d, 2f to be really stable – new domain boundaries appeared neither after half a year storing the crystal at room temperature, nor after heating it up to 100°C. Only above ~100°C we observed pairs of domain boundaries to be created one after another, until all the surface was covered by a dense pattern of boundaries.

The crystallographic rearrangement in LSCO well explains all the memory effects in resistivity. In fact, as the $b$-axis switches along or transverse to the current, following the direction of the magnetic field, the sample exhibits just the $b$-axis or the $a$-axis resistivity (if the axis aligning is perfect). Using resistivity as a probe, we can examine the kinetics of the domain-boundary motion *in situ* at various temperatures – the importance of *in-situ* measurement is that it allows us to confirm the genuine magnetic-field effect by virtue of avoiding any accidental strain as a possible cause of structural changes. Figure 3 displays the resistivity variation upon slowly rotating the magnetic field at 265.5 K. Interestingly, a clear delay in the resistivity switching (*i.e.* $\Delta\rho_{ab}$ is zero at $\alpha \approx 30°$ (60°) for the clockwise (counterclockwise) rotation rather than at 45°) turns out to be independent of the rotation speed in the range 2–6°/min; this indicates that the domain motion is rather quick and goes faster than we rotate the field. This delay in switching comes from the simple fact that the magnetic field becomes capable of swapping the crystal axes only when its projections onto the axes differ by more than ≈ 5 T, comparable to the apparent threshold field in Fig. 1c. We have obtained quite similar data upon rotating the field at temperatures from 190 to 300 K, and found that even at the lowest temperature the domain-boundary motion is still too fast to show up in the dependence on the rotation rate.

To date, the only known compounds that exhibit similar phenomenon have been the ferromagnetic shape-memory alloys, where the magnetic field acting on the ferromagnetic moment of a sample produces a torque sufficient for swapping the crystallographic axes[1,2]. Such considerable magneto-crystal effects remain exotic even among ferromagnets, which powerfully interact with the field. Nevertheless, the present experiments explicitly demonstrate that the magnetic manipulation of the crystal structure is possible in virtually non-magnetic LSCO, however counterintuitive this may sound. Our intuition is, in fact, misled by the ostensible importance of a strong magnetism: A ferromagnetic moment is indeed strongly affected by the magnetic field; however, if spins are not coupled to the lattice, they easily rotate with field, and the effect of the magnetic field is never delivered to the lattice. In reality, it is the magnetocrystalline *anisotropy* that allows the torque to be generated, and provides an energy gain for the crystal rearrangement. The magnetic anisotropy is caused by the tendency of the lattice to align spins to a preferred direction, and may have similar scale in both ferro- and antiferromagnets, even though the former are many thousand times more magnetic. The magnetic anisotropy is actually often employed for aligning non-ferromagnetic powders suspended in liquid and for texturing materials upon solidification[10] – in these cases, just a minute torque tells the system which direction to adopt. What is peculiar in LSCO is that the torque appears to be capable of rotating the crystal axes in a *solid*.

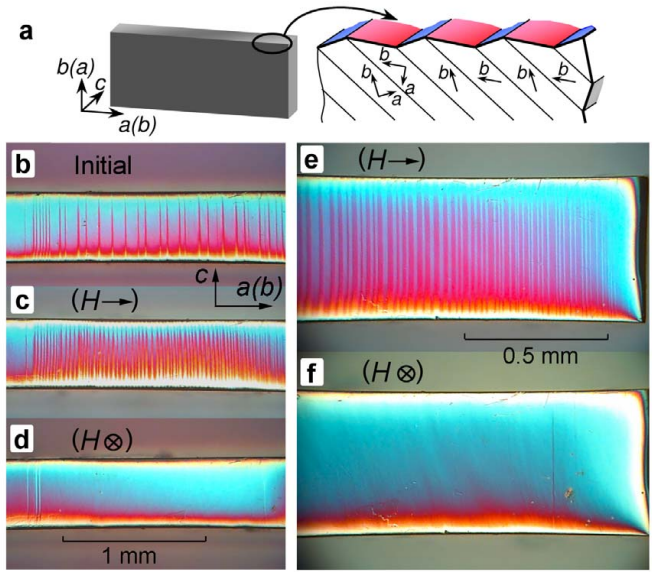

**Figure 2.** Effect of a strong magnetic field on the crystal domain structure, observed with a polarized-light microscope at room temperature. **a**, The orthorhombic-twin pattern, emerging in LSCO below the T-O transition, induces on polished $ac/bc$ faces of crystals a washboard-like modulation which can be easily observed with an optical microscope[8]. **b-f**, Polarized-light images of the same crystal taken *before* and *after* application of the 14-T magnetic field along the directions indicated in panels. The vertical stripes of predominantly blue (red) color on the crystal surface correspond to the $ac$ ($bc$) domains. Initially, the crystal was twinned with larger fraction of $ac$ (blue) domains (**b**). An application of the 14-T magnetic field parallel to the plane of observation changed the domain structure, apparently aligning the $b$-axis along the field direction and thus favoring the $bc$ (red) domains (**c**); a magnified view of the right-hand-side end of the same crystal is shown in **e**. A subsequent application of the magnetic field perpendicular to the plane of observation favored the $ac$ (blue) domains (**d**, **f**). Note that the magnetic field is capable of completely removing the twin boundaries and converting the crystal into the single-domain state (**f**).

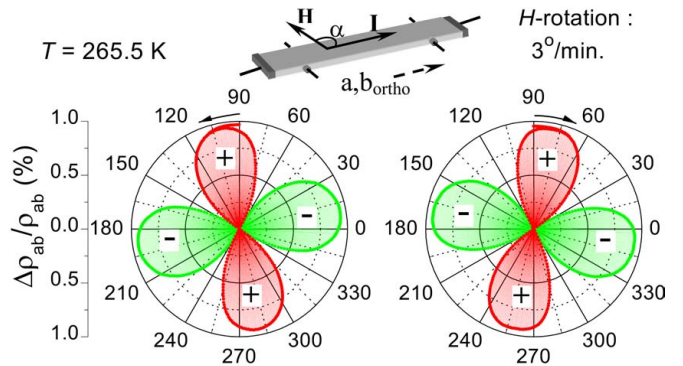

**Figure 3.** Variation of the resistivity, $\Delta\rho_{ab}/\rho_{ab}$, at 265.5 K upon slow rotation of the 14-T magnetic field within the $ab$ plane (the definition of angle $\alpha$ is shown in the inset). The magnetic field was first applied at $\alpha \approx 90°$, and then the data were taken during continuous field rotation at a rate of 3°/min counterclockwise, left panel, and clockwise, right panel. The red (green) segments in the polar plots indicate positive (negative) $\Delta\rho_{ab}$. During the measurements, the crystallographic $b$-axis – the direction of better conduction – switches successively along and transverse to the sample's axis, following the field rotation, which generates the resistivity modulation; note a clear delay (10-15°) in the resistivity switching with respect to the field rotation. A pinning of the domain boundaries smooths the axes rotation, thus resulting in the continuous variation of $\Delta\rho_{ab}/\rho_{ab}$.

The driving force for the magnetic shape-memory phenomena in LSCO is its remarkable in-plane magnetic anisotropy[4]: the susceptibility along the orthorhombic *a* and *b* axes in lightly doped LSCO indeed differs by up to several times (Fig. 4). Moreover, the susceptibility anisotropy is not restricted to the antiferromagnetically ordered state, but persists far above the Néel temperature[4], which allows the magnetic shape-memory effects to occur in both antiferromagnetic (including undoped $La_2CuO_4$) and paramagnetic

LSCO crystals. Such a pronounced magnetic anisotropy being brought about by just a minor lattice distortion points to an unusually strong coupling of the spins with the crystal lattice, where spins, owing to the spin-orbit coupling, tend to follow the orientation of copper-oxygen bonds[11]. Figure 4 further illustrates how the crystal rearranges, trying to gain the magnetic energy; namely, when the magnetic field is applied along the low-susceptibility direction (*a*-axis), the crystal swaps the axes orientation so that the susceptibility along the field direction increases. The energy gain ($\Delta M \cdot H$) is, however, fairly small, only up to $5 \times 10^4$ erg/cm³; thus, given that the LSCO crystal spends this energy for swapping its axes, a force that the crystal can generate upon expanding in one direction by $(b - a) / a \approx 1\%$ would not exceed $\approx 50$ N per 1-cm² area – about an order of magnitude less than what *ferromagnetic* shape-memory alloys[2,12] can generate. The fact that we still observe the axes rotation in LSCO crystals points to a high mobility of their crystallographic domain boundaries, which readily move even under relatively weak force. It is interesting to note that, owing to the high mobility of domain boundaries and to the spin alignment along the *b*-axis[4] (inset in Fig. 4), we can also observe a "reverse" magnetic shape-memory effect: The orientation of all spins in a LSCO crystal can be switched just by tightly squeezing the crystal with fingertips.

What is the potential impact of the discovered magnetic shape-memory effects on the material science? First of all, they visually demonstrate a strong coupling between the magnetic subsystem and the lattice in copper oxides, which people thought they knew more or less everything about, thus offering a missing link in our understanding of these compounds. Also, they expand the general view on what the magnetic field can possibly do on non-ferromagnetic materials. Since the essential origin of these curious effects is the magnetic anisotropy, which is rather abundant in materials, similar phenomena may turn out to be fairly common in other compounds as well; it may just because people have not considered their existence that they have been overlooked.

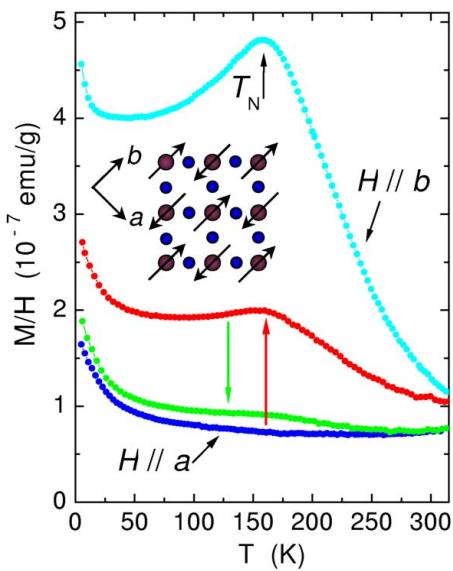

**Figure 4.** In-plane anisotropy in the magnetic susceptibility of LSCO (*x* = 0.01) that makes the crystal-axes alignment with the magnetic field possible. The susceptibility along the *a*-axis (blue curve) and *b*-axis (cyan curve) was measured at *H* = 0.5 T in a mechanically 100%-detwinned LSCO crystal (the peak in the cyan curve corresponds to the Néel transition). Measurements with the "*H // a*" geometry were repeated after application of a 15-T field at room temperature along (red) and then transverse to (green) the initial direction of the *a*-axis. With the strong magnetic field applied along the low-susceptibility direction, the crystal tends to switch the orientation of axes in order to gain magnetic energy; similarly to the situation in Figs. 2e, 2f, the axis alignment is almost perfect for one field direction but only partial for the other, probably because of some uniaxial strain caused by crystal defects. Inset shows the schematic spin arrangement in the Cu (brown) - O (blue) plane; because of a weak out-of-plane spin canting[3,4,11] in LSCO, the susceptibility along the spin easy axis is larger than the transverse one[4].

1. Murray, S. J. *et al.* Large field induced strain in single crystalline Ni–Mn–Ga ferromagnetic shape memory alloy *J. Appl. Phys.* **87,** 4712–4717 (2000).

2. O'Handley, R. C., Murray, S. J., Marioni, M., Nembach, H. & Allen, S. M. Phenomenology of giant magnetic-field-induced strain in ferromagnetic shape-memory materials. *J. Appl. Phys.* **87,** 5774–5776 (2000).

3. Kastner, M. A., Birgeneau, R. J., Shirane, G. & Endoh, Y. Magnetic, transport, and optical properties of monolayer copper oxides. *Rev. Mod. Phys.* **70,** 897–928 (1998).

4. Lavrov, A. N., Ando, Y., Komiya, S. & Tsukada, I. Unusual magnetic susceptibility anisotropy in untwinned $La_{2-x}Sr_xCuO_4$ single crystals in the lightly-doped region. *Phys. Rev. Lett.* **87,** 017007-1–017007-4 (2001).

5. Borsa, F. *et al.* Staggered magnetization in $La_{2-x}Sr_xCuO_4$ from $^{139}$La NQR and $\mu$SR: Effects of Sr doping in the range $0 < x < 0.02$. *Phys. Rev. B* **52,** 7334–7345 (1995).

6. Ando, Y., Lavrov, A. N., Komiya, S., Segawa, K. & Sun, X. F. Mobility of the doped holes and the antiferromagnetic correlations in underdoped high-$T_c$ cuprates. *Phys. Rev. Lett.* **87,** 017001-1–017001-4 (2001).

7. Tranquada, J. M., Sternlieb, B. J., Axe, J. D., Nakamura, Y. & Uchida, S. Evidence for stripe correlations of spins and holes in copper oxide superconductors. *Nature* **375,** 561–563 (1995).

8. Ando, Y., Segawa, K., Komiya, S. & Lavrov, A. N. Electrical resistivity anisotropy from self-organized one-dimensionality in high-temperature superconductors. *Phys. Rev. Lett.* **88,** 137005-1–137005-4 (2002).

9. Ando, Y., Lavrov, A. N. & Segawa, K. Magnetoresistance anomalies in antiferromagnetic $YBa_2Cu_3O_{6+x}$: fingerprints of charged stripes. *Phys. Rev. Lett.* **83,** 2813–2816 (1999).

10. de Rango, P. *et al.* Texturing of magnetic materials at high temperatures by solidification in a magnetic field. *Nature* **349,** 770–772 (1991).

11. Thio T. *et al.* Antisymmetric exchange and its influence on the magnetic structure and conductivity of $La_2CuO_4$. *Phys. Rev. B* **38,** 905–908 (1988).

12. Ullakko, K., Huang, J. K., Kantner, C., O'Handley, R. C. & Kokorin, V. V. Large magnetic-field-induced strains in $Ni_2MnGa$ single crystals. *Appl. Phys. Lett.* **69,** 1966–1968 (1996).

### Acknowledgements

We thank K. Segawa for invaluable technical assistance, and V. M. Loktev for discussions.

Correspondence and requests for materials should be addressed to Y. A. (e-mail: ando@criepi.denken.or.jp).